# Jupyter Notebooks on GitHub: Characteristics and Code Clones

Malin Källén[a], Ulf Sigvardsson[b], and Tobias Wrigstad[a]

a   Uppsala University
b   work done while at Uppsala University

**Abstract**   Jupyter notebooks has emerged as a standard tool for data science programming. Programs in Jupyter notebooks are different from typical programs as they are constructed by a collection of code snippets interleaved with text and visualisation. This allows interactive exploration and snippets may be executed in different order which may give rise to different results due to side-effects between snippets.

Previous studies have shown the presence of considerable code duplication — code clones — in sources of traditional programs, in both so-called systems programming languages and so-called scripting languages.

In this paper we present the first large-scale study of code cloning in Jupyter notebooks. We analyse a corpus of 2.7 million Jupyter notebooks hosted on GitHub, representing 37 million individual snippets and 227 million lines of code. We study clones at the level of individual snippets, and study the extent to which snippets are recurring across multiple notebooks. We study both identical clones and approximate clones and conduct a small-scale ocular inspection of the most common clones.

We find that code cloning is common in Jupyter notebooks – more than 70 % of all code snippets are exact copies of other snippets (with possible differences in white spaces), and around 50 % of all notebooks do not have any unique snippet, but consists solely of snippets that are also found elsewhere. In notebooks written in Python, at least 80 % of all snippets are approximate clones and the prevalence of code cloning is higher in Python than in other languages.

We further find that clones between different repositories are far more common than clones within the same repository. However, the most common individual repository from which a Jupyter notebook contains clones is the repository in which itself resides.

# The Art, Science, and Engineering of Programming

## 1 Introduction

Data science, that is processing, analysing and extracting knowledge from large quantities of data, has emerged as a new inter-disciplinary field or new research paradigm, and also an increasingly important component in industry, as many companies strive to be "data-driven". The emergence and rapid growth of this field is fuelled by the availability and easy access to vast quantities of data, the relative ease with which such data sets can be gathered with new technology, and the availability of easy-to-use computational tools that hide most of the complicated data crunching and computation behind (relatively speaking) easy interfaces. This allows a new class of programmers — that would not traditionally view themselves as such — to explore data sets and use statistical methods for business decisions, in research, and in society.

"New-born data scientist programmers" typically have a strong background in maths. While there are similarities between maths and programming, such as the importance of abstractions, the imperative nature of many computational workflows are fundamentally different than mathematics. As researchers and societal organisations increasingly rely on data science, curating the "programs" that generate the plots etc. used to support decisions becomes important for reproducibility, traceability and accountability. This typically falls under software engineering, which further departs from mathematics or most subjects taught outside of a computer science curriculum.

To support the use of data science to produce valid results across a wide range of fields, it is important to study the programs involved to understand challenges and needs. For example, the prevalent use of untyped programming languages like Python and R lends a certain flexibility to programming not possible in e. g., Haskell and Scala, but at the same time may cause unpredicted behaviour due to type violations that might be hard to detect if the shape of the output falls within what is expected.

As a first step of a longer journey towards understanding and supporting programming in data science, we study code written using Jupyter notebooks, a popular tool for data scientists. These are computational notebooks that mix text, graphics and code to develop a "computational narrative" that can be used to document both a particular (executable) method and its rationale in conjunction with the results. In order to build an understanding of "data science code", we have obtained more than 2.7 million Jupyter notebooks, mined from GitHub. The rationale for choosing Jupyter notebooks are several: it supports a wide variety of programming languages (including the aforementioned), a large corpus of programs can be relatively easily mined from GitHub, and Jupyter notebooks have already been studied to some extent in prior work, albeit at a smaller scale.

In prior work from 2018, Rule, Tabard, and Hollan studied Jupyter using a 600 GB data set with over 1 million notebooks. To promote further studies, Rule, Tabard, and Hollan curated a smaller "sample data set" at 5 GB with about 6000 randomly selected notebooks from 1000 repositories [12]. As pointed out by Lopes, Maj, Martins, Saini, Yang, Zitny, Sajnani, and Vitek, creating such a data set is complicated due to the "non-trivial amount of duplication in source code". For example, the method of selection can greatly impact the result. For example, the 10K Python projects on GitHub with the *most stars* contain 28 % duplication whereas the 10K Python projects with the *most commits* have 44 % duplication. (For comparison, less than 25 % of all non-small Python files hosted on GitHub are unique.) Without understanding how code is cloned in Jupyter notebooks, it is hard to trust any results obtained by studying Rule, Tabard, and Hollan's sampler data set.

In order to build a quantitative understanding of Jupyter notebooks at scale, we must understand code cloning in notebook code. Identification of code clones can provide important information for API developers. For example, code blocks that are frequently cloned may be candidates for new functions, which would decrease the overall software maintenance burden. In particular, defects detected in those blocks may affect many different code bases and its developers may never get to know about the defects and be able to fix them. If, instead, the code containing the defect is wrapped in a function, the fix will automatically propagate to all code bases using that function once the single block of code was updated.

Furthermore, knowledge about cloning is also important for researchers in software engineering. As pointed out by Lopes, Maj, Martins, Saini, Yang, Zitny, Sajnani, and Vitek, a high prevalence of clones in code bases used as study objects may severely skew the results of research. Not taking this into consideration may lead to incorrect conclusions in scientific studies and misdirected tool development and understanding of the nature of code. This makes studies of online repositories such as GitHub particulariy important, since source code used in scientific studies is often collected from there.

In this paper, we extend the scholarly understanding of Jupyter notebooks and in particular how code is cloned across notebooks. While code cloning studies on Python and other languages often focus on clones of entire files or individual functions, we study clones at the level of notebook cells. We make the following contributions:

- To understand Jupyter notebooks, we report on a number of statistics from our large corpus, including number of code cells, notebook sizes, and language distributions, and relate these numbers to results from previous studies (section 4).
- We present the first large-scale study of code cloning in Jupyter notebooks. We study clones at the level of individual snippets, and study the extent to which snippets are recurring across multiple notebooks. We study clones at two different levels of similarity (section 5) and conduct a small-scale ocular inspection of the most common clones (appendix C).
- In the spirit of repeatability, we make our scripts and analysis programs, and our full data set of close to 1 TB data, available (appendix D).

**Outline** In addition to the above, section 2 briefly introduces Jupyter notebooks and code clones; section 3 discusses related work, and provides a comprehensive overview of most large-scale code cloning studies to date, ranging from the scale of 1500 files to 261 million; and section 7 concludes. There are also four appendices: appendix A discusses data acquisition; appendix B describes how we extracted language information from the notebooks; appendix C contains a qualitative discussion of the contents of the most common clones; and appendix D points to the GitHub repository containing our artefacts.

## 2 Background

**Jupyter Notebooks**  Jupyter notebooks is a tool for the interactive development and presentation of data science projects across dozens of programming languages. Jupyter notebooks projects are organised in notebooks, which are files that consist of a sequence of "cells" that contain code and output of code (if any), or text. Output can be raw data or visualisations. Text can be documentation that constructs a "computational narrative", explanations, equations, etc. Together, the cells make up a kind of literate program, except that cells can be executed individually and in any order.

Every notebook uses a specific kernel, which controls the language that executes the code in all cells of a notebook. The kernel state persists over time and is shared between cells, which means for example that cells executed in different order can change the behaviour — and output — of a document. Internally, Jupyter Notebook files are stored on disk in JSON format.

The Jupyter project was conceived in 2014 as a language agnostic continuation of the Python-specific IPython Notebook which appeared in prototype form in 2010. The Jupyter project is community-driven and developed openly on GitHub. Jupyter notebooks has enjoyed wide adoption in recent years [13, 20], possibly driven by the surge in popularity of the field of data science. As a web-based interface, Jupyter notebooks provides a low-entry bar access to a powerful computational toolbox. That a notebook is not "just a program" in the traditional sense is evident from the ability to convert it e. g., to HTML or a slide deck. On the other hand, the programs in notebooks are not necessarily very complicated: a recent study by Pimentel, Murta, Braganholo, and Freire explores the proliferation of language features in Python-based notebooks and finds that less than two thirds of all notebooks include any form of conditional.

A deeper investigation of how programmers use Jupyter notebooks, especially as tools to build computational narratives, can be found in the paper by Rule, Tabard, and Hollan. Pimentel, Murta, Braganholo, and Freire studies, among other things, the reproducibility of notebooks.

**Code Clones**  Code entities that are similar to each other are referred to as clones. Depending on the level of similarity, clones are categorised into three different types:
- Type 1 clones are code segments that are identical except possible differences in comments and white spaces.
- Type 2 clones are structurally equal, and may have differences in types, identifiers and literals, in addition to the differences allowed for type 1 clones.
- Type 3 clones may have the same differences as type 2 clones. Additionally, a few statements may differ between the code segments. This type of clone is also referred to as near-miss clones.

Some authors also use the term type 4 clones to describe code segments that have the same functionality, but are implemented differently.

While the definitions of type 1 and type 2 clones are clear, that of type 3 clones is relatively vague, and have been interpreted in different ways in the literature. In practice, it is determined by the implementation of the tool that is used for clone detection, and some (normally configurable) threshold value for the similarity.

Code entities that are clones of each other are referred to as a clone group. If the clone group contains only two code entities, it can also be referred to as a clone pair. When the entities in question are entire code files, we talk about file level clones, and when the entities are functions, we call them function level clones. As will be seen in section 3, not all clones are file or function level clones, but they may also be any code sequence with a predefined minimum number of tokens.

The term "clone" is somewhat misleading, since it indicates that one code entity in a clone pair is copied from the other one [5]. This is not necessarily the case, but a clone pair may arise by accident when the same functionality is implemented in two different code entities by different developers or at different points in time [1, 5]. This type of clone is referred to as accidental clones by Al-Ekram, Kapser, Holt, and Godfrey.

## 3 Related Work

**Jupyter Notebooks**   One of the earliest studies of Jupyter notebooks is from Rule, Tabard, and Hollan [13]. To study the narrative of computational notebooks, they downloaded the 1 227 573 Jupyter notebooks publicly available on GitHub in July 2017. They found that among the 97.8 % of the notebooks that contained code, the median number of lines of code is 85, and the maximum value is >400 000. According to their analysis, 82 % of their data set is written in Python and around 1 % in R and Julia respectively. 14.9 % of the notebooks do not specify a programming language. From the most common words in the descriptions of the repositories from where the notebooks were taken, the authors concluded that the most common usage of Jupyter notebooks is for education and machine learning.

Studying a sample of the notebooks used in scientific projects, Rule, Tabard, and Hollan [13] concluded that Jupyter notebooks are used for iterative analyses, more often than for presenting rich narratives. They interviewed 15 academic data scientists focusing on notebooks in research (not education) and found Jupyter notebooks to be mainly used for experimenting, although a few interviewees also used them to communicate with non-programmers. At the same time, the design of Jupyter notebooks seeks to facilitate documentation and communication of research. According to the authors, there is a tension between different fields of application: Experimenting often results in messy notebooks, and in order to use them for documentation or communication, notebooks must be cleaned, which tend to never happen.

A tension between different usages of notebooks was also found by Kery, Radensky, Arya, John, and Myers [3] in an interview study with data scientists. This study concludes that data scientists *do* create narratives in their notebooks — by (re)organising code cells rather than through extensive explanations in text cells.

In 2019, Pimentel, Murta, Braganholo, and Freire [10] studied Jupyter notebooks, focusing on quality and reproducibility. They analysed all non-empty, unique, valid notebooks in GitHub repositories (excluding forks) in April 2018 — a total of 1 159 166 notebooks. They report that the language is undefined in 2.98 % of their notebooks, while 93.32 %, 1.31 % and 0.93 % of the notebooks are written in Python, R and Julia respectively. The authors looked at the 863 878 notebooks with an explicit Python version and unambiguous execution order. Of these, 24.11 % executed without error and only 4.03 % produced the same result as the one that was stored in the notebook.

Another paper with focus on the quality of Jupyter notebooks is written by Wang, Li, and Zeller [20]. They analysed 1982 Jupyter notebooks written in Python using the official Python style guide (PEP8) checker. This experiment reported more than one error in every three lines of code. This can be compared to less than one error in every seven lines in the ordinary Python scripts located in the same projects as the notebooks.

**Code Cloning**   The prevalence of code clones in open source software has been studied in several earlier studies. Here, we summarise the code cloning studies most relevant to this paper. These were largely made in the last 10 years, due to the rise of large code repositories (e. g., GitHub) and availability and maturity of clone detection tools.

The closest work to ours is by Koenzen, Ernst, and Storey [4], to our knowledge the only other study of code clones in Jupyter notebooks. They designate theirs an exploratory study that does not aim for generalisable results. They studied cloned code cells in a small sample from the corpus of Rule, Tabard, and Hollan consisting of 897 projects. These projects contain 6386 notebooks, which in turn contain 78 020 code cells. From this sample, they excluded projects that contain less than 28 code cells in total. (They do not specify the number of such projects.) The study deals with intra-repository clones only and the authors found that on average 7.6 % of the code cells in a repository are type 3 clones of each other. The minimum, median and maximum clone frequencies in the repositories are 0.0 %, 5.0 % and 47.5 % respectively. Approximately 1/4 of the clones are type 1 clones. Studying a sample of 500 clones, the authors concluded that the most common main activities in the code cells are (in order) visualisation, machine learning and function definitions. They also performed a small user study, where they found that the participants frequently clone code found online or in internal sources when possible.

Yang, Martins, Saini, and Lopes [21] studied function-level clones in 909 288 GitHub projects (30 986 363 files) written in Python. They found that 86 % of the functions in the corpus are exact copies of the remaining 14 %. Using type 3 clones, allowing clones to differ up to 20 %, they found that 88 % of the functions are clones of the remaining 12 %.

An impressive study by Lopes, Maj, Martins, Saini, Yang, Zitny, Sajnani, and Vitek [7] deals with cloning at file level in all Python, Java, C++ and JavaScript projects that resided on GitHub at the time of their study, and were not forks of other projects. Of the projects included, 893 197 (31 602 780 files) are written in Python, 1 481 468 (72 880 615 files) are written in Java and 364 155 (61 647 575 files) are written in C++. In the Python projects, they found that only 29 % of the files are unique, with the remaining 71 % being exact copies of these. For Java, C++ and JavaScript the portion of unique files is 60 %, 27 % and 6 % respectively. The most duplicated file is the empty file and the second most duplicated file contains one empty line only.

Lopes, Maj, Martins, Saini, Yang, Zitny, Sajnani, and Vitek picked all files with a unique set of tokens (modulo comments) to study the prevalence of type 3 clones, using `SourcererCC` [17] with an 80 % similarity threshold. Note that this is a subset of the unique files, composed of 8 620 326, 40 786 858, 14 425 319, and 13 587 850 files written in Python, Java, C++ and JavaScript respectively. This means that type 3 clones that are also type 1 clones are excluded from the results, and so are clones that only differ in the token separators used or the order of the tokens. However, the results are reported in terms of fractions of the whole projects. They found that 9 % of the Python files are near-miss clones of each other. The corresponding numbers for Java, C++ and JavaScript are 26 %, 10 % and 2 % respectively. A manual inspection of a sample of the files indicated that a majority of the type 3 file level clones of Python are auto generated code for initialisation of a Django application.

Code cloning in Java code on file level was also studied by Ossher, Sajnani, and Lopes [9], who considered 13 241 projects, constituting 3 237 910 files, mainly downloaded from SourceForge, Apache, Java.net and Google Code. They found that 5.2 % of the files are exact copies of each other. Using different (very conservative) estimates, they found that the clone ratio, involving near-miss clones, is between 10 % to 15 %. The most frequently cloned type of file in their study seems to be third party library files. The authors could not find any relation between the project size and the clone frequency, but a comparison with a study made by Mockus [8] indicates that cloning is less frequent in Java projects than in e.g. C projects.

A recent study on code cloning in Java projects was done by Gharehyazie, Ray, Keshani, Zavosht, Heydarnoori, and Filkov [2]. They studied the presence of type 1 and type 2 clones in 8599 GitHub projects written in Java. To be taken into consideration in the clone analysis, a code sequence had to consist of at least 20, 30 and 50 tokens respectively. Out of the 8599 projects, 5753 projects (1 040 000 files) contain clones, and between 5 and 10 % of the code basis in these projects consists of cloned code. Most clone pairs occur within projects. Likewise, clones between projects in the same domain seemed to be more common than clones between projects of different domains.

Roy and Cordy [11] studied within-project function clones in a number of open source projects written in C, Java and C# respectively. Their corpus consists of 10 C, 7 Java and 6 C# projects, containing 10 877, 1519 and 5495 files respectively. In order to exclude empty functions, they only included the functions that were at least 3 lines in pretty printed format. Just as the results by Ossher, Sajnani, and Lopes, the results of this study indicate that the clone characteristics of a project are independent of the project size, but that there is a distinct difference between projects written in different languages. The authors found that cloning is most common in the Java projects, where 7.2 % and 14.4 % of the lines of code are involved in type 1 and type 3 clones (with at most 20 % difference) respectively. The C projects contain the least amount of cloned code with 1.0 % of the lines of code being involved in type 1 clones, and 8.4 % of the lines of code being involved in type 3 clones.

Koschke and Bazrafshan [6] studied the prevalence of cloned code sequences in C and C++ code, using the Ubuntu C sources together with the data sets provided by two other authors, in total 7844 projects (826 560 files), excluding duplicates.

**Jupyter Notebooks on GitHub: Characteristics and Code Clones**

Contrary to the authors of the studies summarised above, they extracted commonly used library code from the projects into which it was copied, and considered each library (once) as a project of its own instead considering every copy of it. They found that around 2.5 %, 1.9 % and 1.2 % of the code base consists of type 1 clones of at least 30, 50 and 100 token respectively. The corresponding numbers for type 2 clones are 40 %, 28 % and 16 % respectively, including all clones detected by their automatic tool. They also used two different filters to exclude accidental clones. (Accidental clones are described by Al-Ekram, Kapser, Holt, and Godfrey [1]. The concept is also discussed by Koschke [5], although he does not use the term "accidental clones".) Then, the type 2 clone rates are 22 % to 23 %, 18 % to 19 % and 11 % to 12 % respectively. This reduction of the clone rate is in line with the results of Al-Ekram, Kapser, Holt, and Godfrey, who came to the conclusion that accidental clones constitute a substantial part of all clones. In accordance with other studies that compare different languages, Koschke and Bazrafshan found that the clone characteristics are language dependent: the C++ code has a higher clone frequency than the C code.

We have made an attempt to summarise the studies described in this section in table 1, but the studies are hard to compare since they are performed differently and present different data. It is clear from the table that the studies cover clones on different levels and with different minimum sizes. Likewise, Koenzen, Ernst, and Storey [4] and Roy and Cordy [11] only consider within-project clones while the other studies also cover inter-project clones. Moreover, while some studies use their whole corpora when looking for type 2 and type 3 clones, Gharehyazie, Ray, Keshani, Zavosht, Heydarnoori, and Filkov [2] do not include projects without clones and Lopes, Maj, Martins, Saini, Yang, Zitny, Sajnani, and Vitek exclude all files that have the same token hashes (which includes all type 1 clones). Another important difference is that the studies use different tools, which use different algorithms. In particular, since there is no generally accepted definition of type 3 clones, in practice, the different tools use different definitions of type 3/near-miss clones, even when they use the same threshold value. There are also other aspects of the methodologies that differ. For example, Koschke and Bazrafshan [6] do not consider extensively copied third party library files, which may be expected to make a big difference, not at least since Ossher, Sajnani, and Lopes conclude that external library files are the most commonly cloned type of files. Last, the results are presented in different ways. While Yang, Martins, Saini, and Lopes [21] report the fraction of functions that are clones of each other, Roy and Cordy [11] report the fraction of lines involved in function clones. More importantly, the results of Yang, Martins, Saini, and Lopes and the type 1 clone results of Lopes, Maj, Martins, Saini, Yang, Zitny, Sajnani, and Vitek are presented differently than the other results presented in table 1: The former report how many code units (files) are unique, as opposed to code units that are copies of these. The other results are reported as the percentage of code units that are involved in clones, or the mean code frequency for repositories. Hence, note that the columns "Unique" and "Clones" in table 1 are not the opposite of each other!

That said, all studies indicate that there is a substantial amount of code cloning present in (open source) C, C++, C# Java, JavaScript and Python code. Moreover, the studies seem to agree that the extent of code cloning is language dependent.

▪ **Table 1** Summary of code cloning studies covered in section 3 (Code Cloning). Studies higher up in the table are more comparable to ours. Note that unique is not the opposite of clones, but involves one copy from each clone group.

| Study | Year | Projects | Files | Inter-proj | Language | Level | Type | Min size | Unique | Clones |
|---|---|---|---|---|---|---|---|---|---|---|
| [4] | 2020 | <897 | 6386 | no | multiple | snippet | 3 | 0 | ? | 7.6 %[1] |
| [21] | 2017 | 909 288 | 30 986 363 | yes | Python | function | 1 | 0 | 14 % | ? |
| [21] | 2017 | 909 288 | 30 986 363 | yes | Python | function | 3 | 0 | 12 % | ? |
| [7] | 2017 | 893 197 | 31 602 780 | yes | Python | file | 1 | 0 | 29 % | ? |
| [7] | 2017 | 1 481 468 | 72 880 615 | yes | Java | file | 1 | 0 | 60 % | ? |
| [7] | 2017 | 364 155 | 61 647 575 | yes | C++ | file | 1 | 0 | 27 % | ? |
| [7] | 2017 | 1 755 618 | 261 676 091 | yes | JavaScript | file | 1 | 0 | 6 % | ? |
| [7] | 2017 | 1 481 468 | 72 880 615 | yes | Python | file | 3 | 0 | ? | 9 %[2] |
| [7] | 2017 | 364 155 | 61 647 575 | yes | Java | file | 3 | 0 | ? | 26 %[2] |
| [7] | 2017 | 1 755 618 | 261 676 091 | yes | C++ | file | 3 | 0 | ? | 10 %[2] |
| [7] | 2017 | 1 481 468 | 72 880 615 | yes | JavaScript | file | 3 | 0 | ? | 2 %[2] |
| [9] | 2011 | 13 241 | 3 237 910 | yes | Java | file | 1 | 0 | ? | 5 % |
| [9] | 2011 | 13 241 | 3 237 910 | yes | Java | file | 3 | 0 | ? | 10–15 %[3] |
| [2] | 2019 | 5753 | 1 040 000 | yes | Java | token | 2 | 20/30/50 t | ? | 5–10 %[4] |
| [11] | 2009 | 10 | 10 877 | no | C | function | 1 | 3 loc | ? | 1 %[5] |
| [11] | 2009 | 7 | 1591 | no | Java | function | 1 | 3 loc | ? | 7 %[5] |
| [11] | 2009 | 6 | 5495 | no | C# | function | 1 | 3 loc | ? | 6 %[5] |
| [11] | 2009 | 10 | 10 877 | no | C | function | 3 | 3 loc | ? | 11 %[5] |
| [11] | 2009 | 7 | 1591 | no | Java | function | 3 | 3 loc | ? | 20 %[5] |
| [11] | 2009 | 6 | 5495 | no | C# | function | 3 | 3 loc | ? | 13 %[5] |
| [6] | 2016 | 7844 | 826 560 | yes | C/C++ | token | 1 | 30 t | ? | 3 % |
| [6] | 2016 | 7844 | 826 560 | yes | C/C++ | token | 1 | 50 t | ? | 2 % |
| [6] | 2016 | 7844 | 826 560 | yes | C/C++ | token | 1 | 100 t | ? | 1 % |
| [6] | 2016 | 7844 | 826 560 | yes | C/C++ | token | 2 | 30 t | ? | 40 % |
| [6] | 2016 | 7844 | 826 560 | yes | C/C++ | token | 2 | 50 t | ? | 28 % |
| [6] | 2016 | 7844 | 826 560 | yes | C/C++ | token | 2 | 100 t | ? | 16 % |
| [6] | 2016 | 7844 | 826 560 | yes | C/C++ | token | 2 | 30 t | ? | 22–23 %[6] |
| [6] | 2016 | 7844 | 826 560 | yes | C/C++ | token | 2 | 50 t | ? | 18–19 %[6] |
| [6] | 2016 | 7844 | 826 560 | yes | C/C++ | token | 2 | 100 t | ? | 11–12 %[6] |

[1] Clone frequency is the mean for all repositories  [2] (A super set of) type 1 clones excluded  [3] Conservative estimate  [4] Projects without clones excluded  [5] Percentage of lines of code  [6] Excluding accidental clones

## 4 The Corpus

Note: for brevity, data acquisition is relegated to appendix A. Below we report on the size and language distribution of the data set.

### 4.1 Size of the data set

The number of downloaded notebooks is 2 739 464. Of these, a small number, 6088, come from projects that are forks of other project, and are excluded from the analysis below. Moreover, from 13 998 notebooks, we could not extract any information, typically because they do not contain JSON data or the JSON data was ill-formed. Some

**Jupyter Notebooks on GitHub: Characteristics and Code Clones**

■ **Table 2** Sizes of notebooks in the corpus

|  | Bytes | Code cells | Non-empty LOC | LOC, incl. empty |
| --- | ---: | ---: | ---: | ---: |
| Min | 3 | 0 | 0 | 0 |
| 10th percentile | 2468 | 2 | 5 | 5 |
| 25th percentile | 5296 | 4 | 19 | 22 |
| Median | 13 185 | 9 | 47 | 53 |
| Mean | 363 680.75 | 13.60 | 83.38 | 96.50 |
| 75th percentile | 31 945 | 17 | 102 | 118 |
| 90th percentile | 57 939 | 29 | 196 | 229 |
| Max | 108 484 960 | 1433 | 462 116 | 462 118 |

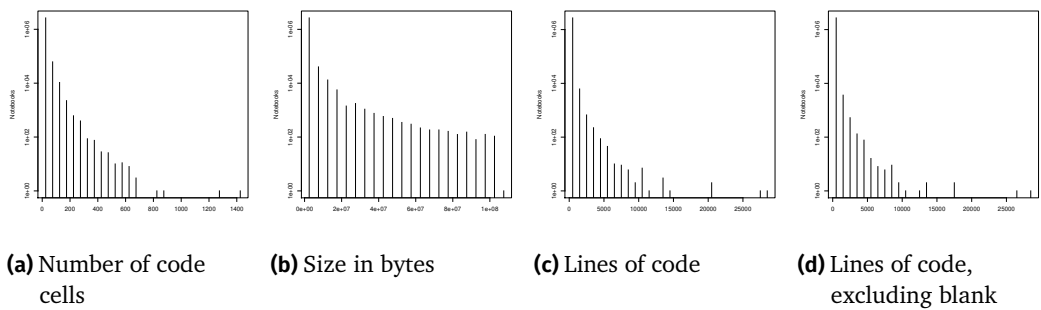

**(a)** Number of code cells    **(b)** Size in bytes    **(c)** Lines of code    **(d)** Lines of code, excluding blank

■ **Figure 1** Distribution of notebook sizes. Note the logarithmic scale of the y axes. Zero counts are omitted in the plots. One outlier is excluded from figure 1c and figure 1d.

of these were due to our inability to process GIT LFS storage. These notebooks are also excluded from the analysis.

During the analysis of the remaining 2 719 378 notebooks, there were two notebooks for which we could not extract any code related information due to format errors. Data for these are reported as if there was no code cells at all. Also due to format errors, there was one notebook from which we could not extract line counts or code. For that notebook, we report data as if the code cells do not contain any code.

The total size of the analysed notebooks is approximately 921 GB (including text cells and metadata). In total, these notebooks contain 36 974 714 code cells. From now on, we refer to the code stored in a code cell as a snippet. The snippets of the analysed notebooks comprise 226 744 094 lines, excluding blank lines but including comments. With blank lines included, the total number of lines is 262 414 943. The maximum, minimum, mean and median values for the notebook sizes can be found in table 2, together with the 10th, 25th, 75th and 90th percentiles.

Figure 1 shows the corresponding distribution of sizes, with log scales on the y axes. However, one outlier, the maximum number of lines of code (with blank lines included and excluded: 462 118 and 462 116 respectively) has been excluded from the plots. Also, zero counts are not plotted. As can be seen in figure 1 as well as in table 2, the distributions of the notebook sizes are highly right-skewed.

### 4.2 Programming Language Distribution in Jupyter Notebooks on GitHub

Jupyter notebooks support several different programming languages. To be able to select notebooks based on languages, and to interpret our results and compare to other studies, we set off to analyse which languages are actually used in our corpus. This turns out to not be entirely straightforward. A detailed description of the methodology for determining languages in a notebook can be found in appendix B. The results are presented below.

Using the method described in appendix B, we have been able to identify the language in slightly more notebooks than Pimentel, Murta, Braganholo, and Freire and considerably more notebooks than Rule, Tabard, and Hollan. One reason for this may be that recently uploaded notebooks specify the language more often than older ones. Another reason is the fact that we are looking for language values in several different fields. Both Rule, Tabard, and Hollan and Pimentel, Murta, Braganholo, and Freire seem to have extracted language information from metadata.language_info.name. When that field lacked a value, Rule, Tabard, and Hollan also checked metadata.language.

In our corpus, 93.07 % of the notebooks specify the language in metadata.language_info.code.name and 0.12 % specify the language in metadata.language but not metadata.language_info.name. This means that using the methodology of Rule, Tabard, and Hollan and Pimentel, Murta, Braganholo, and Freire, we would have failed to identify the languages in 6.81 % and 6.93 % of the notebooks respectively (compared to the actual ratio, 2.19 %). Nevertheless, our results are consistent in that Python is by far the most common programming language used in Jupyter notebooks uploaded on GitHub.

**Table 3** Language distribution in the 2 719 378 notebooks

| Language | Notebooks | Percent |
|---|---|---|
| Python | 2 592 892 | 95.35 % |
| Julia | 22 336 | 0.8214 % |
| R | 21 432 | 0.7881 % |
| Scala | 5155 | 0.1896 % |
| Other, known | 18 099 | 0.6656 % |
| Unknown | 59 464 | 2.187 % |

There are 2083 notebooks that have language information stored in more than one of the fields listed in appendix B, and with differing information in the fields. This is interesting, but the field chosen for language identification should not bias the results since 2083 notebooks is only 0.77 ‰ of the whole corpus.

## 5 Code Clones in Jupyter Notebooks on GitHub

We have analysed snippets that are clones of each other, i. e., "snippet-level clones." The analysis consisted of two steps: First, we studied snippets that are identical modulo differences in white space. Because far from all clones are identical copies of other code segments (see e. g., table 1), we have also studied near-miss clones. The steps are described in section 5.1 and section 5.2 respectively. To analyse identical snippets, we used our own program NotebookAnalyzer, specifically written for this task, together with some post processing scripts. To analyse near-miss clones, we used the tool SourcererCC by Sajnani, Saini, Svajlenko, Roy, and Lopes [17], our own program SccOutputAnalyzer

**Jupyter Notebooks on GitHub: Characteristics and Code Clones**

and a number of post processing scripts. `NotebookAnalyzer`, `SccOutputAnalyzer` and all post processing scripts are available online, see appendix D.

### 5.1 Copy Modulo Whitespace Clones — CMW

As mentioned in section 2, type 1 clones are code entities that are identical modulo white spaces and comments. Since we analyse snippets written in several different languages, removing all comments is unfeasible, and we have chosen to analyse a type of clone that we will refer to as CMW clones, where "CMW" is an abbreviation of "Copy Modulo White space". This clone type differs from type 1 clones only in the fact that we include comments in the analysis. Hence, two snippets that have identical code but differences in comments are not CMW clones, while snippets that have differences only in white spaces are. In section 5.1, clones and uniqueness refers to CMW clones and CMW uniqueness.

#### 5.1.1 Methodology

As a first step, we removed all white spaces (including newlines) from the snippets. Consequently, in this section "empty snippet" may refer to a snippet containing white spaces but nothing else. Next, we computed the MD5 hash of each snippet. In order to identify clones, we compared the resulting hashes: snippets with identical hashes are considered being CMW clones of each other. There is a risk that the same MD5 sum was obtained for some snippets that are not identical, but thanks to our large data set, these are unlikely to skew the results. We also computed the median number of lines of code (rounded down when the median is not an integer), including comments, of each clone group. As line breaks are removed before hashing, the line count can differ between the snippets in a clone group. When reporting line counts of clones below, we use the median line count of the clone group.

Using CMW clone data, we identified cloned notebooks. We consider two notebooks $N_1$ and $N_2$ to be clones if both notebooks have $n$ code cells, and for $i = [1, n]$, the $i$:th snippet in $N_1$ is a CMW clone of the $i$:th snippet in $N_2$. Thus, only the code cells in the notebooks matter in notebook clone identification.

Thereafter, we computed the clone frequency for each notebook. Here, the clone frequency refers to the fraction of the snippets of the notebook that are clones of other snippets (as opposed to unique in the corpus). We examined if there is any correlation between the size of a notebook (measured in number of snippets) and its clone frequency. Since neither the number of code cells nor the clone frequency is normally distributed, we used a Spearman's rank correlation test, using the R function `cor.test` with `method="spearman"` and `alternative="two.sided"`. Next, we wanted to see if the clone frequencies are different for different languages. We tried an ordinary ANOVA, but the residuals appeared not to be normally distributed. Hence, we used the non-parametric equivalent, Kruskal-Wallis ANOVA, by calling the R function `kruskal.test`. Since the notebooks with the language value UNDEFINED can be written in any language, we excluded them from this analysis. The Kruskal-Wallis test indicated a significant difference between the languages. Therefore, we made a post hoc analysis applying pairwise Wilcoxon rank sum tests using the R function `pairwise.wilcox.test`. To control our family-wise type I error rate, we used Hochberg p-value adjustment (which has a higher power than Bonferroni p-value adjustment). Hence, we gave the parameters p.adjust.method="hochberg", paired=FALSE, and alternative="two.sided" to pairwise.wilcox.test.

Last, we counted the number of connections for each notebook as follows: Let $G$ be a graph where each node represents a notebook in our corpus. For each pair of snippets that are clones of each other, there is an edge connecting the notebooks containing the two snippets. The number of connections of a notebook is simply the number of edges in $G$ attached to the node of that notebook. The normalised number of connections is the number of connections for a notebook, divided by the number of snippets in the notebook (or 0 for notebooks without any snippets). We refer to connections between notebooks in the same repository as intra connections and to connections between notebooks in different repositories as inter connections.

To determine whether notebooks are connected to other notebooks in the same repository to a larger, or smaller, extent than to notebooks of other repositories, we used the following method: For each notebook $N$ in $G$, we counted the number of connections $C_i, i = [1, m]$ to notebooks in each repository $R_i, i = [1, m]$, excluding repositories that contain no notebooks connected to $N$, and the one where $N$ resides. Let $C_0$ represent the number of intra connections for a notebook. Having computed all values of $C_i$, we computed the mean number of inter connections $IC$ for $N$ as the mean of $C_i, i = [1, m]$ (or 0 if there are no inter connections for the notebook). Thereafter, we performed a Wilcoxon signed rank test on all values of $C_0$ and $IC$ using the R function wilcox.test with parameters alternative="two.sided" and paired=TRUE. We also compared $C_0$ to the total number of inter connections, $SC$ (that is the sum of all connection counts $C_i, i = [1, m]$) for each notebook the same way.

### 5.1.2 Results and Discussion

The most common clone is the empty snippet, which occurs 2 914 109 times. This is in line with the results of Lopes, Maj, Martins, Saini, Yang, Zitny, Sajnani, and Vitek, which show that an empty file is the most common file in their corpus, followed by a file containing one empty line. Accordingly, almost 8 % of all snippets are empty (or contain only white spaces). This clone is most likely accidental in most of the cases, and provide little interesting information about the code. Therefore, empty snippets are excluded from the analysis below.

**Descriptive Statistics** The total number of snippets that have at least one clone is 24 773 774, clustered in 4 462 974 clone groups. The number of unique snippets in the corpus is 9 286 828,[1] which gives a clone ratio of 72.73 %. All clones in the corpus can be found in at least 2 different notebooks. However, 324 740 (11.94 %) of the notebooks have at least one clone from itself, that is some snippet occurs at least 2 times in the same notebook.

---

[1] While the total number of code cells (reported in section 4.1) is 36 974 714, the total number of snippets is 36 974 711 as there is one notebook, containing 3 code cells, from which we could extract the number of code cells, but due to a format error, we could not extract the code. Hence, NotebookAnalyzer reported no snippets for that notebook.

**Jupyter Notebooks on GitHub: Characteristics and Code Clones**

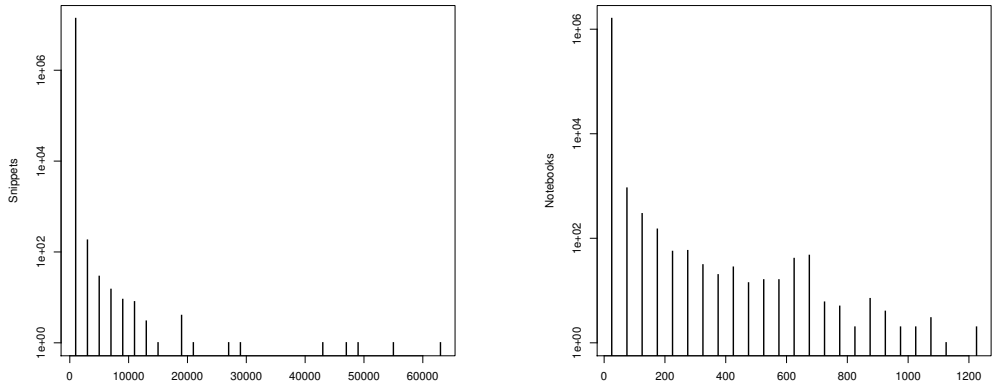

**(a)** Distribution of number of *snippet* clone occurrences

**(b)** Distribution of number of *notebook* clone occurrences

■ **Figure 2** Note the logarithmic scale of the y axes. Zero counts are omitted in the plots.

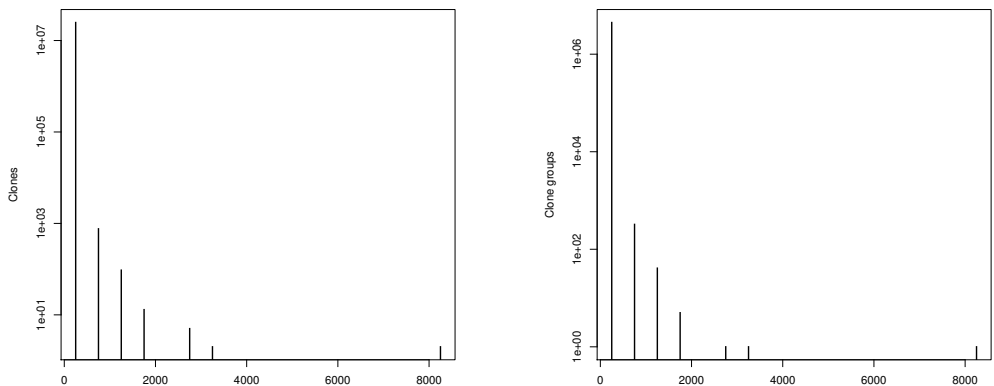

**(a)** Clone sizes

**(b)** Clone group sizes

■ **Figure 3** Distribution of clone sizes and clone group sizes, measured as the median LOC of the snippets. Note the logarithmic scale of the y axis. Zero counts are omitted in the plot.

Figure 2a shows the distribution of the sizes of the clone groups, with log scale on the y axis. The distribution is highly right-skewed and most snippets occur less than 200 times.

Figure 3a and figure 3b show the distribution of snippet size, measured as number of lines of code, among the clones[2] and among the clone groups respectively. These distributions are highly right skewed. In other words, small clones are considerably

---

[2] Recall: we use the median LOC of the clone group as LOC value for all clones in that group.

▪ **Table 4** Line counts for CMW clones and clone groups (columns 2–3) and CMW clone frequencies (column 4)

|  | Clone sizes | Clone group sizes | CMW clone frequencies |
| --- | --- | --- | --- |
| Min | 1 | 1 | 0.00 % |
| 10th percentile | 1 | 1 | 0.00 % |
| 25th percentile | 1 | 1 | 18.18 % |
| Median | 2 | 3 | 98.91 % |
| Mean | 5.83 | 7.48 | 64.53 % |
| 75th percentile | 6 | 8 | 100.00 % |
| 90th percentile | 14 | 17 | 100.00 % |
| Max | 8318 | 8318 | 100.00 % |

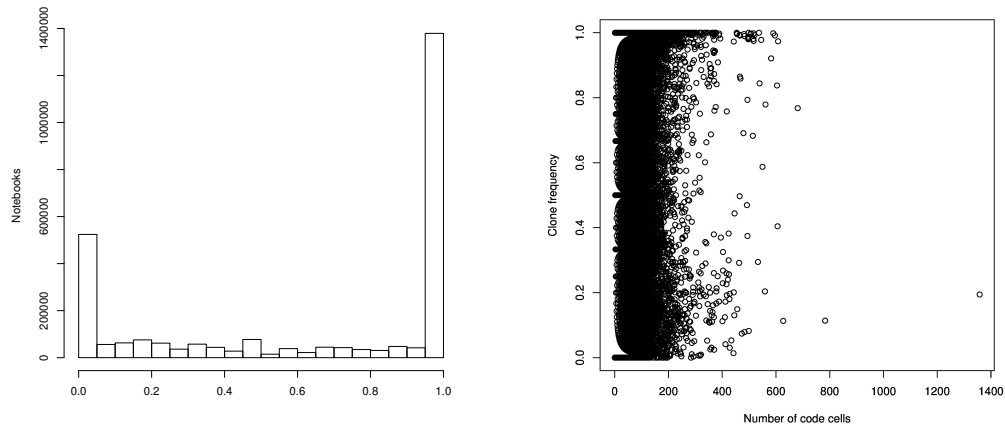

**(a)** Distribution of CMW clone frequencies.

**(b)** CMW clone frequencies vs. number of code cells in notebooks.

▪ **Figure 4**

more common than large clones, and the probability that a snippet will be cloned is higher if it is small.

The maximum, minimum, mean and median values for the line count of the clones and clone groups, together with the 10th, 25th, 75th and 90th percentiles, are presented in table 4. Here, it is clear that most clones as well as clone groups have very few lines of code. Moreover, the mean and percentile values are smaller for the clones than for the clone group, which indicates that clone groups that contain small snippets are larger than those that contain large snippets.

Out of the notebooks, 1 359 173 (49.98 %) contain only cloned snippets, while 343 341 (12.63 %) contain only unique snippets.

**Cloned Notebooks** The distribution of the notebook clone occurrences is presented in figure 2b. Like the distribution of snippet occurrences, it is highly right skewed. In total, 1 292 653 notebooks have a clone.

**Jupyter Notebooks on GitHub: Characteristics and Code Clones**

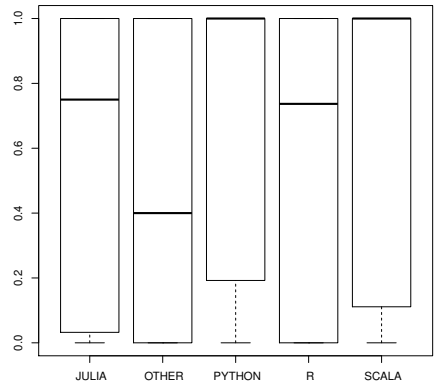

**Figure 5** Clone frequencies for different languages

**Table 5** Adjusted p-values from post hoc comparison of clone frequencies across languages

|       | Python | Julia | R | Scala |
|-------|--------|-------|---|-------|
| Julia | $<2.22 \times 10^{-16}$ | | | |
| R     | $<2.2 \times 10^{-16}$ | 0.0052 | | |
| Scala | 0.0031 | $2.7 \times 10^{-13}$ | $<2.2 \times 10^{-16}$ | |
| Other | $<2.2 \times 10^{-16}$ | $<2.2 \times 10^{-16}$ | $<2.2 \times 10^{-16}$ | $<2.2 \times 10^{-16}$ |

**Clone Frequencies** The distribution of the clone frequencies is presented in figure 4a. As can be seen, the clone frequency is often either very low (<5 %) or very high (>95 %). High clone frequencies are more common than the low frequencies.

The maximum, minimum, mean and median values for the clone frequencies can be found in table 4, together with the 10$^{th}$, 25$^{th}$, 75$^{th}$ and 90$^{th}$ percentiles. Again, we see that high clone frequencies are much more common than low clone frequencies.

Figure 4b shows how clone frequencies vary with the number of code cells in the notebooks. We see no indication of a correlation between the two characteristics. However, the result of the Spearman correlation test indicates a significant but relatively weak correlation: $\rho = 0.2383, p < 2.2 \times 10^{-16}$.

Figure 5 shows a box plot of the clone frequencies for the different languages. The Kruskal-Wallis ANOVA indicates a significant difference in clone frequencies between the different languages: $\chi^2 = 3748, p < 2.2 \times 10^{-16}$, which is consistent with earlier studies discussed in section 3. The post hoc analysis suggests a statistically significant difference for all combinations of languages. (Recall that notebooks with unknown languages are excluded from this analysis.) See table 5 for p values. In all cases but two (Python-Scala and Julia-R), the difference is significant on the 0.001 level. The difference between Python and Scala as well as between Julia and R is significant on the 0.01 level. In figure 5, it seems like the clone frequency is highest for Python, followed by Scala, while the clone frequency is lowest for the (heterogeneous) group of other languages.

Malin Källén, Ulf Sigvardsson, and Tobias Wrigstad

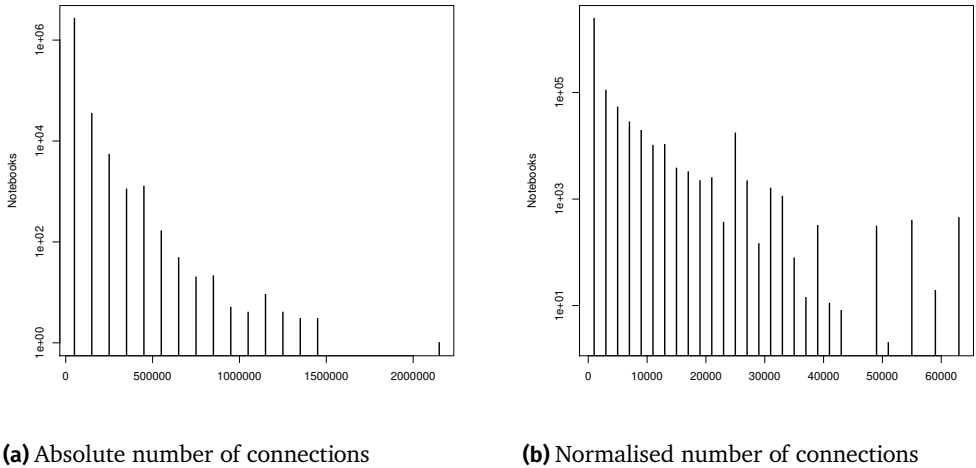

**(a)** Absolute number of connections

**(b)** Normalised number of connections

**Figure 6** Distribution of the number of connections for each notebook. Note the logarithmic scale on the y axes. Zero counts are omitted.

**Table 6** Notebook connections via CMW clones

|  | Absolute number | Normalised number |
|---|---|---|
| Min | 0 | 0.00 |
| 10$^{th}$ percentile | 0 | 0.00 |
| 25$^{th}$ percentile | 3 | 0.90 |
| Median | 108 | 11.60 |
| Mean | 9615.30 | 920.59 |
| 75$^{th}$ percentile | 4173 | 351.13 |
| 90$^{th}$ percentile | 30 199 | 1922.89 |
| Max | 2 170 917 | 63 048.00 |

**Connections** Figure 6 shows the distribution of the number of connections for a notebook, with logarithmic scale on the y axes. Table 6 shows the corresponding mean, min, max and percentile values. Both the absolute and the normalised number of connections are clearly right-skewed. The maximum number of connections is 2 170 917, which is almost as large as the total number of notebooks in the corpus. Note that there may be many connections between the same notebook pair, and a notebook may also be connected to itself several times, so the corresponding notebook is not necessarily connected to 2 170 917 *different* notebooks. There is one notebook in which the *average* snippet is connected to over 63 000 other snippets.

The results from the Wilcoxon signed rank test indicate a significant difference between the number of intra connections ($C_0$ described in section 5.1.1) and the mean number of inter connections ($IC$ described in section 5.1.1): $V = 1.665 \times 10^{12}$, $p < 2.2 \times 10^{-16}$. Likewise, they indicate a statistically significant difference between $C_0$ and the total number of inter connections ($SC$, also described in section 5.1.1) for each notebook: $V = 1.608 \times 10^{11}$, $p < 2.2 \times 10^{-16}$.

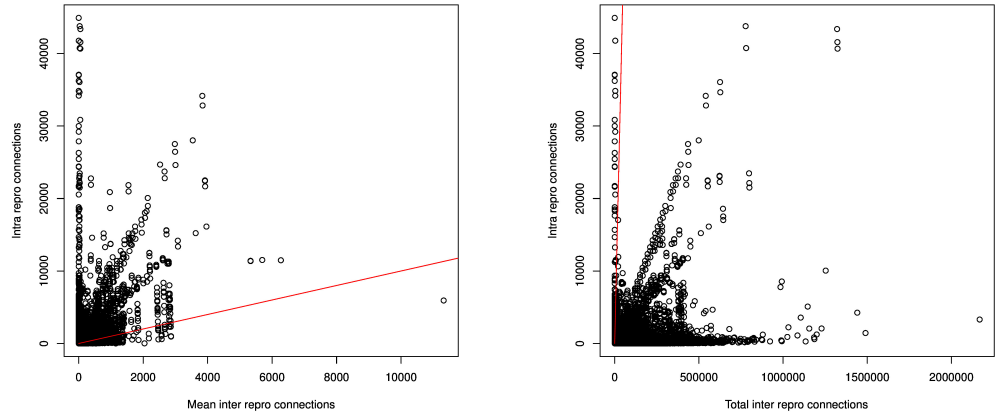

**Figure 7** Number of intra repository connections compared to mean number (left) or total number (right) of inter repository connections for each notebook

Figure 7 plots $C_0$ against $IC$ (left), and $C_0$ is against $SC$ (right). The red lines show where $C_0 = IC$ and $C_0 = SC$ respectively. From these pictures, it is clear that $C_0$ is larger than $IC$, but smaller than $SC$. In other words, if a notebook $N$ is connected to another notebook $NC$, the probability that $NC$ resides in the same repository as $N$ is significantly higher than the probability that it resides in a particular other repository that $N$ is connected to. On the other hand, the probability that $NC$ resides in the same repository as $N$ is significantly lower than the probability that it lives in *any* other repository. The latter conclusion may be an explanation for the low clone prevalence found by Koenzen, Ernst, and Storey.

## 5.2 Near-Miss Clones

To identify near-miss clones, we used the clone detector SourcererCC [17] (commit `afc5755273fc91754ee08326a6946fad85e85b94` cloned from GitHub [14]). This tool removes comments before the clone analysis, and therefore cannot analyse clones where comments are denoted differently. Hence we had to choose one language to do the near-miss clone analysis of the notebook snippets. Since more than 95 % of the notebooks in our corpus are written in Python, we decided to analyse snippets from these.

### 5.2.1 Methodology

SourcererCC's analysis runs in two steps: First it tokenizes the code on file or block level and thereafter it identifies clones using the tokenized data. Since we wanted to analyse clones on snippet level, we dumped the source code from each code cell found in a notebook for which Python was identified as a language to a separate file. For more information about the code used for this, see appendix D. Thereafter, we tokenized each such file using the SourcererCC file level tokenizer. We specified `#` as language primitive for inline comments (`comment_inline` in `config.ini`) and `"""` as opening and close tag for block comments (comment_open_tag and comment_close_tag in config.ini). For separators (in config.ini), we used the default value.

Next, we ran the clone detector. To include all snippets in the analysis, we set the minimum token value for a clone (MIN_TOKENS in sourcerer-cc.properties) to 0 and the maximum token value (MAX_TOKENS in sourcerer-cc.properties) to 500 000 000. However, it appeared as SourcererCC ignores clones containing 0 or 1 token [15]. We kept the default threshold value of 8, meaning clones will be flagged at 80 % similarity.

The execution of SourcererCC, which went on for more than 6 weeks on a 128 core server, was interrupted twice because the server went down. Both times, we managed to resume. During the processing of the tokenized input data, with some interval, SourcererCC writes the number of the last processed line to a recovery file. When resuming an earlier execution, it starts reading from the line following the last line registered in the recovery file. This means that some clone pairs will be reported twice. On the other hand, it guarantees that a line that is not properly processed will be reread and processed after the recovery. To avoid double reporting of clones, we identified all lines in the clone pair file that had a duplicate, and removed the duplicate before continuing with the post processing described below.

We also found lines containing binary data in the clone pair files, which we assume was caused by the sudden interruptions of the execution. Accordingly, we removed everything that was not a number, a ',' or a line feed (which are the expected characters in the clone pair file). Since any line that is only partly processed will be processed again after the recovery, this shouldn't cause any data loss. We also found five lines on the wrong format (that is more than four numbers separated by commas). We cannot explain this, and have reported it to the SourcererCC developers [16]. We excluded these lines from the analysis. Since the total number of clone pairs reported by SourcererCC is >18 000 000 000, this should not cause any bias in the results.

Just as in section 5.1, we exclude empty snippets from the analysis.

In the clone analysis, SourcererCC excludes blank lines and comments. Hence, in the discussion of near-miss clones, we will use the number of source lines of code reported by SourcererCC as line count. This means that comments and blank lines are not included. Likewise, snippets only containing comments are considered empty.

Based on the output data from SourcererCC, we identified cloned and unique snippets in the corpus and computed clone frequencies and connection counts using SccOutputAnalyzer. We also performed the same statistical analyses of clone frequency and connections as we did for the CMW clones. A reference to the code can be found in appendix D.

### 5.2.2 Results and Discussion

Out of the 35 450 312 Python snippets in our corpus, 3 673 406 (10.36 %) contain 0 lines of code. These are excluded from the analysis in this section.

**Descriptive Statistics**   The total number of Python snippets that have at least one clone (according to SourcererCC) is 25 339 340, which corresponds to 79.74 % of the analysed snippets. This is higher than the numbers reported in section 5.1.2, which could be expected since the near-miss clones of a code base are a super set of the

**Jupyter Notebooks on GitHub: Characteristics and Code Clones**

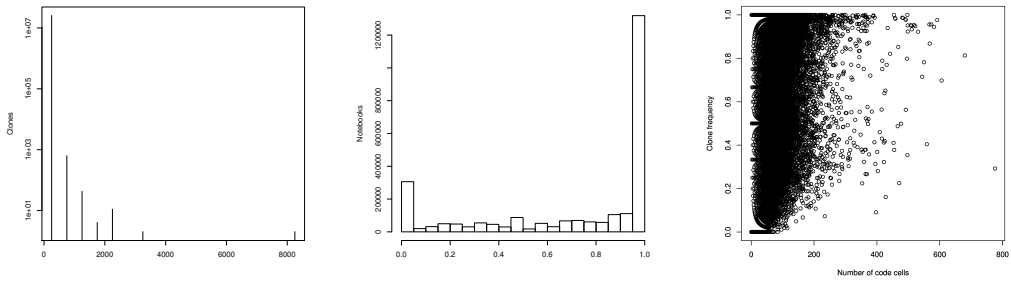

**(a)** Distribution of line counts for near-miss clones. Zero counts are omitted.

**(b)** Distribution of near-miss clone frequencies for Python notebooks.

**(c)** Near-miss clone frequencies vs. number of code cells in Python notebooks.

■ **Figure 8**

■ **Table 7** Line counts for near-miss clones

| Min | 10$^{th}$ perc. | 25$^{th}$ perc. | Median | Mean | 75$^{th}$ perc. | 90$^{th}$ perc. | Max |
|---|---|---|---|---|---|---|---|
| 1 | 1 | 1 | 2 | 5.53 | 6 | 13 | 8318 |

CMW clones. Moreover, Python notebooks have the highest clone frequencies in our corpus.

Out of the 2 592 892 Python notebooks, 744 548 contain at least one near-miss clone from itself. This corresponds to 28.71 %, which is more than two times the value reported for CMW clones. In other words, a little bit more than half of the notebooks that clone snippets from itself do not contain any CMW clone from itself. (Remember that CMW clones are a subset of the near-miss clones.)

Out of the Python notebooks, 1 258 545 ones (48.54 %) consist of near-miss clones only, while 140 215 (5.41 %) only contain unique snippets. Not surprisingly, the last number is smaller than the one reported for CMW clones. However, the fraction of the notebooks that contain only cloned snippets is similar. This indicates that notebooks that contain near-miss clones but not CMW clones in general also contain unique snippets.

Figure 8a shows the distribution of the sizes of the near-miss clones, measured as the number of lines of source code. The distribution is similar to that of CMW clones (figure 3a). The maximum, minimum, mean and percentile values for the line counts of the near-miss clones can be found in table 7. Just as the CMW clones, most near-miss clones are very small.

**Clone Frequencies**   The distribution of frequencies of near-miss clones in Python notebooks are presented in figure 8b. The most salient difference from the CMW clone frequency distribution (figure 4a) is that fewer notebooks have a really low clone frequency (<5 %). On the contrary, frequencies over 95 % are by far the most common for both clone types. Minimum, median, maximum and mean values for the near-miss clone frequency in Python notebooks are presented in table 8, together with the 10$^{th}$, 25$^{th}$, 75$^{th}$ and 90$^{th}$ percentiles. Once more, we see that the clone frequencies

▪ **Table 8** Near-miss clone frequencies for Python notebooks

| Min | 10th perc. | 25th perc. | Median | Mean | 75th perc. | 90th perc. | Max |
|---|---|---|---|---|---|---|---|
| 0.00 % | 0.00 % | 50.00 % | 96.00 % | 72.88 % | 100.00 % | 100.00 % | 100.00 % |

▪ **Table 9** Notebook connections via near-miss clones

|  | Absolute number | Normalised number |
|---|---|---|
| Min | 0 | 0.00 |
| 10th percentile | 0 | 0.00 |
| 25th percentile | 18 | 3.00 |
| Median | 632 | 70.00 |
| Mean | 13 988.89 | 1502.15 |
| 75th percentile | 12 459 | 1000.39 |
| 90th percentile | 53 277 | 4056.55 |
| Max | 1 679 144 | 80 520.00 |

are generally high. Naturally, low clone frequencies are less common for near-miss clones than for CMW clones (table 4).

In figure 8c, we see how the near miss clone frequency varies with the number of code cells. We can see a small tendency towards a positive correlation between the two quantities, and indeed, the Spearman correlation test indicates a highly significant, although relatively weak correlation: $\rho = 0.1233, p < 2.2 \times 10^{-16}$. This correlation is weaker than for the CMW clones. In other words, the results suggest that the CMW clone frequency grows a little bit faster than the near-miss clone frequency when the size of the notebook increases.

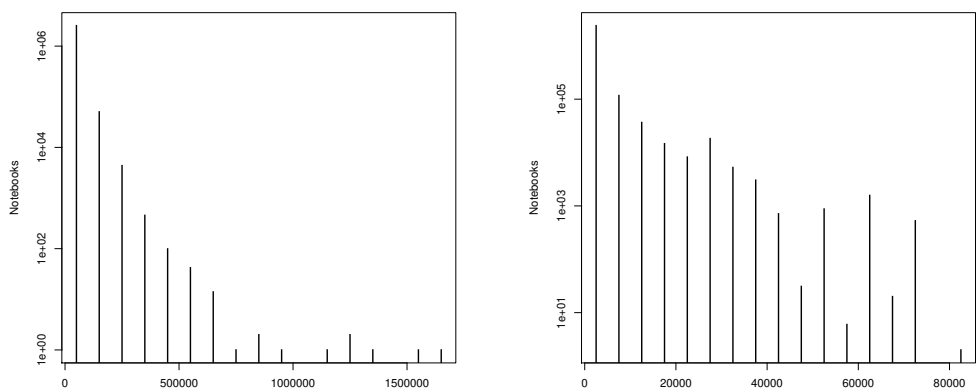

**(a)** Absolute number of connections    **(b)** Normalised number of connections

▪ **Figure 9** Distribution of the number of connections between near-miss clones for each notebook. Logarithmic scale on the y axes. Zero counts are omitted in the plots.

**Jupyter Notebooks on GitHub: Characteristics and Code Clones**

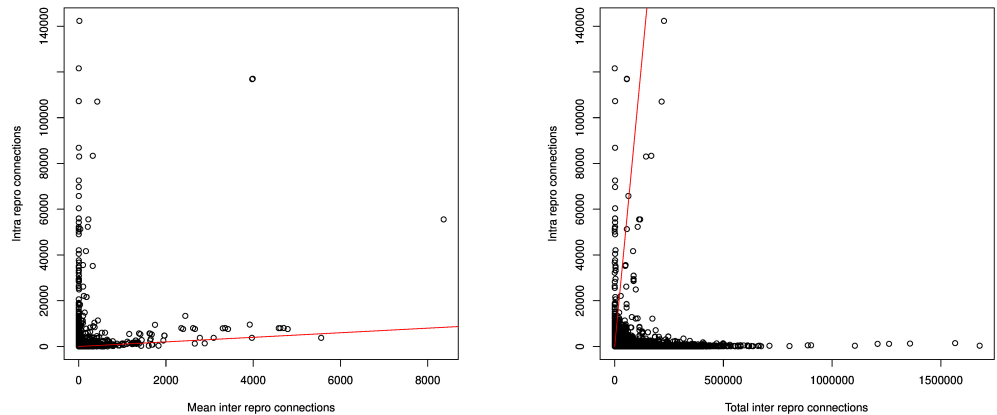

■ **Figure 10** Number of intra repository connections compared to mean number (left) or total number (right) of inter repository connections for each notebook

**Connections**  In figure 9, we present the distributions of connections between the Python notebooks. In accordance with what can be seen in figure 6, the distributions are highly right-skewed, in particular the absolute number of connections. Table 9 shows the percentiles together with min, max, median and mean values of the connections. Here, we see a higher number of connections than in table 6, reflecting the fact that the number of near-miss clone instances is higher than the number of CMW clone instances. The only exception is the max value, which is higher for CMW clones. This is discussed in section 6.

The signed Wilcoxon rank sum tests indicate a statistically significant difference between the number of intra repository connections ($C_0$) and the mean number of inter connections ($IC$): $V = 2.050 \times 10^{12}, p < 2.2 \times 10^{-16}$, and between $C_0$ and the total number of inter repository connections ($SC$): $V = 9.158 \times 10^{10}, p < 2.2 \times 10^{-16}$.

From figure 10, we conclude that $C_0$ is larger than $IC$ and smaller than $SC$. The red lines show where $C_0 = IC$ and $C_0 = SC$ respectively. This means that, just as for the CMW clones, most near-miss clones are inter-project clones, but the single repository with which a notebook shares most clones is its own repository.

## 6 Discussion

**Why Do Programmers Clone Code?**  In prior work on code cloning, clone rates are high across all studied languages. It is interesting to ponder *why* this is so. Concurrent with this work, Koenzen, Ernst, and Storey [4] conducted a small-scale study with eight participants that were given a number of simple tasks to complete in Jupyter notebooks, and studied how the subjects solved their problems. Half of the participants worked by copying and pasting their own code from previous solutions, when problems were similar. All participants used the web extensively to solve their problems, consulting among other things API documentation and Stack Overflow. Our own informal expe-rience is that this tendency to program by "copying-and-patching" existing code is prevalent among both students and professionals, as it quickly provides a program that can be run to inspect results and behaviour and compare with expectations.

Studying non-notebook Python code, Lopes, Maj, Martins, Saini, Yang, Zitny, Sajnani, and Vitek [7] found that popular web frameworks like Django, which use static code generation of boilerplate code to simplify programming tasks, drive up the code clone amount. In Jupyter notebooks, no such frameworks exist (to the best of our knowledge). We also find that the clone frequency is lower than in non-notebook Python code.

Many short snippets are likely accidental, which is also discussed in appendix C. Many of them consist of single variable names used to render textual representation of data in the notebooks. We see a tradition of naming variables consistently across notebooks, e. g., many data frame variables are named df. We also see many recurring imports or groups of imports of libraries commonly used together. Neither of these clones are particularly interesting, nor do they incur much technical debt.

Although we may speculate, it is hard to make any conclusive statement about why clone rates are high, and why this is seemingly the norm for software development. Some aspects of the data set might drive the number up; we discuss this further down under Threats to Validity.

We are puzzled by the clone frequencies being highest for Python *and* Scala. Both are object-oriented (or multi-paradigm) class-based languages, and object-orientation is typically considered good for creating reusable components, but as we have stated—we see little evidence of reusable components (like functions) being defined in the notebooks. One possible cause for the Python result is found in its design philosophy which explicitly states that there should be "one—and preferably only one—obvious way to do it." This might mean that Python code is more susceptible to accidental clones, but we stress that this is highly speculative, and certainly does not explain the Scala results; especially since, in our subjective view, Python and Scala are quite different languages, that should appeal to different user bases. Why these particular languages see more clones than say R or Julia is perplexing. It would be interesting to study this as part of future work.

**Clones at Snippet-Level vs. File- and Function-Level** We study clones at a snippet-level granularity, as opposed to file-level or function-level. Conceptually, snippets and functions are similar, especially as "parameters" can be transferred through shared variables in a global namespace—but whether programmers exploit this in practice or not, we do not know for sure. Despite studying clones at a different granularity, the high frequency of exact copies compared to near-miss clones is on pair with the earlier mentioned studies of clones in Python code [7, 21]. These studies show a higher prevalence of type 1 clones than of type 3 clones that are not type 1 clones. This suggests to future researchers that wish to study Jupyter notebooks—or other Python code—that it is possible to get comprehensive results by considering type 1 clones only. This is an important result, since it is much easier to detect type 1 clones than type 3 clones. The study by Lopes, Maj, Martins, Saini, Yang, Zitny, Sajnani, and Vitek suggests that type 1 clones are most common also for the other languages studied (Java, C++ and JavaScript). However, other studies [6, 11] suggest that type 3 clones that are not type 1 clones are more common than type 1 clones. Hence, one should be cautious to exclude near-miss clones from all code cloning studies.

Almost 50 % of the notebooks in our corpus are clones of each other, that is, they contain the exact same code snippets in the same order as one or more other notebooks. The largest clusters of identical notebooks contain 1000 to 1200 notebooks. Ocular inspection of these reveal that many are instructions from tutorials or course material. In the future, we would like to exclude all such notebooks from the corpus, both those that are seemingly unchanged copies of instructions and those that stem from such instructions but have deviated.

**CMW Clone Connections vs. Near-Miss Clone Connections**   The savvy reader may remark that the maximum number of CMW clone connections — 2 170 917 — is considerably higher than the maximum value of near-miss clone connections. The notebook containing 2 170 917 CMW clone connections is an outlier, which is visible in figure 6. This is a Python notebook and the reason that the number of CMW clone connections is higher is most likely one-token snippets that are common clones. Those do not seem captured by SourcererCC, see section 5.2.1, and are therefore not included in the count of near-miss clone connections. The notebook in question contains 305 code cells, many of which contain just a single variable that is common across the corpus. For example, the snippet a occurs 49 times in this notebook, giving rise to a large number of connections from the notebook to itself as well as to other notebooks.

**Threats to Validity**   Our study is based on Jupyter notebooks hosted publicly on GitHub. The number of notebooks on GitHub have exploded in recent years, but it is not clear how representative these notebooks are for notebooks outside of GitHub. For example, it may be that notebooks that are too sensitive to host publicly or even on GitHub are of a different nature than the ones that are not. We cannot exclude that notebooks hosted on say GitLab are vastly different than those we have studied. It may also be the case that we have missed some search strings when mining GitHub.

We suffered a data loss from a handful of corrupt notebook files. Knowing exactly what files were lost, we wrote a script to download them again, using the same version as in the initial download, but some files were not accessible and could not be downloaded again. Even though the files lost were concentrated to two size classes of notebooks, the relatively small number of files lost (340) should not skew our results. As already discussed in section 5.2.1, we had to clean the output data of SourcererCC's clone analysis. For the reasons mentioned, we do not think that this has biased our results.

A considerable amount of notebooks are seemingly products of tutorials or similar. We have observed recurring instructions in comments and references to course pages etc. If the public notebooks are predominantly developed by students following a small selection of influential tutorials, this will skew our data.

Even though we have seen no such evidence, we cannot competely rule out that many notebooks have been automatically generated from a small number of generators, which would also skew our results.

A couple of SourcererCC idiosyncrasies may have affected our results to a small degree. First, SourcererCC ignores 1-token snippets. Since 1 076 617 of the CMW clones contain less than 2 tokens, the actual number of near-miss clones is most likely even higher than the one reported. Many of the clones excluded are most likely accidental (e. g., a single variable name). Further, SourcererCC only allows the user to specify one set of tags for block comments. We chose """, meaning that "comments" enclosed by ''' are included in the analysis of near-miss clones.

## 7 Conclusions

We have downloaded and studied 2.7 million notebooks from GitHub. This study focuses on basic characteristics of and clones in notebooks. We also plan to continue analysing the corpus in the future, identifying common modules and functions in Jupyter notebooks and look for pitfalls for users of these.

According to measurements in this study, notebooks are in general small (median 13 kB) and contain relatively few code cells (median 9) and few lines of code (median ≈50). However, the distributions are highly right-skewed and some notebooks are very large. In accordance with earlier studies [10, 13], we have found that Python is the dominating language—more than 95 % of the notebooks are written in Python. The second and third most common languages are Julia and R, but their prevalence is less than 1 % each.

The main focus of our study is on code cloning. To our knowledge, at the time of writing, this is the biggest study of clones in Jupyter notebooks, and the only one that deals with inter-project clones. We have seen that code cloning is common in Jupyter notebooks—more than 70 % of all code snippets are exact copies of other snippets (modulo white spaces). Moreover, around 50 % of all notebooks do not have a unique snippet, but consists solely of snippets that are also found elsewhere. In Python notebooks, around 80 % of all snippets are near-miss clones, and this is most likely an under-estimate. The prevalence of cloning is higher in Python than in other languages, but this number should still reflect the whole corpus relatively well since the corpus mainly consists of Python notebooks.

Generally, clones are small (the 90$^{th}$ percentile is 13 to 14 lines of code). The most commonly cloned non-empty snippets contain a single import statement or the name of a single variable. These, as well as many other small clones, are most likely accidental. However, there are snippets that seem to be copied from each other several thousands of times. On the other hand, we have not found any non-empty snippet that occurs in a majority of notebooks.

Studying how clone groups are divided between repositories, we infer that inter-project clones are far more common than intra-project clones. This is opposite to the findings of Gharehyazie, Ray, Keshani, Zavosht, Heydarnoori, and Filkov on Java code. An explanation for this, besides the difference in languages studied, may be that the nature of Jupyter notebooks is different than that of pure source code. Despite the high inter-project clone frequency, our study suggests that the most common individual repository from which a Jupyter notebook contains clones is the repository in which itself resides.

Given the observed [4] tendencies of programmers to solve problems by copy-and-patch from previous solutions to similar problems, we believe code clones will be inevitable until ways are developed that provide an even faster path to a working solution, or tools are developed that are able to successfully fuse clones and near-clones into a single encapsulated unit.

Finally, any study based on code from repositories such as GitHub must account for code clones. This is particularity important when the results are used by developers of tool support for Jupyter notebook code and programmers. Not considering clones could result in wasted or misdirected efforts or development of broken heuristics.

## 8 Acknowledgements

We are indebted to Petr Maj for technical support at many points throughout this project, and to Jan Vitek for many insightful discussions.

## A   Data Acquisition

The corpus of Jupyter notebooks acquired for our study was obtained from GitHub in spring of 2019 for a preliminary study as part of a bachelor thesis project [18]. We used a script developed by Rule, Tabard, and Hollan [13] that uses the GitHub API to query for files with the suffix .ipynb in repositories classified as *Jupyter Notebook* by GitHub. The second constraint excludes IPython repositories (which use an identical file suffix). Notably, the second constraint also causes notebooks in repositories tagged as another language to be missed (which could happen for example when a repository contains multiple files from various languages). In the end, we obtained the URLs to 2 814 381 notebook files (as part of their meta data entries including for example their containing repository, the fork status of this repository etc.) which were downloaded using a custom Python script.[3] Due to limitations enforced by GitHub and our available hardware, the downloading time was eight weeks. As a consequence, files changed name or permissions or were simply removed during the downloading. Empty files were ignored as well as duplicate URLs in the metadata (see [18]). In the end, 2 769 304 individual files were downloaded weighing in at over 900 gigabytes.

When the work for this paper was initiated in the fall of 2019, data corruption of two ZIP archives from the preliminary study caused a loss of notebook files which had to be downloaded again. As the URLs in the meta data included commit data, we were able to attempt to download the identical files as we had initially for the preliminary study. Of the 184 082 corrupted files that had to be downloaded again, we were able to obtain 183 742. In the end, with duplicates removed, our corpus consists of 2 739 464 files.

For a much more detailed breakdown of the data acquisition, see [18]. The URLs and all meta data files for all notebooks are available online [19].

---

[3] Scripts available on GitHub. Script for downloading meta data: http://bit.ly/gh_md_scrape. Script for downloading notebooks: http://bit.ly/nb_download.

## B  How we Identified Programming Languages in Notebooks

The programming language of a notebook can be specified in different ways. We identified the following fields as potential information holders for the programming language:

1. metadata.language_info.name
2. metadata.language
3. metadata.kernelspec.language
4. language inside each code cell

where "." means that the field after "." is stored inside the object preceding the ".". However, every field is not specified in every notebook. In order to identify the language of a notebook, we read these fields in the order they are listed above. As soon as we found a language value, we stopped searching and considered that value being the language of the current notebook. In the cases where we read the language from the code cells, we also verified that the same language is specified in every code cell of the notebook. Thereafter, we categorised the notebooks into 6 different language groups:

- JULIA, if the language value equals "Julia" or "julia"
- PYTHON, if the language value begins with "Python" or "python"
- R, if the language value equals "R" or "r"
- SCALA, if the language value begins with "Scala" or "scala"
- UNDEFINED if the language is not specified in any of the examined fields
- OTHER otherwise

We also identified all notebooks in which the language is specified in more than one field, and the values specified are different.

## C  A Qualitative Review of the Most Common Clones

With the result of our analysis, we ocularly inspected the most common clones. This is sensible from the perspective of CMW clones, where all clones are perfect copies (modulo white spaces), but not from the perspective of near-miss clones, as the clone relation is not transitive. Below, we only discuss CMW clones.

**Clones of At Least 4 Lines**  Looking at snippets of at least 4 lines, many of the most common clones come from various templates and assignments. The most common clone (4846 occurrences) is part of an assignment on neural networks. A large number of clones come from the same course (cs231, about 1400 to 3800 occurrences each), and contain the comment "DO NOT MODIFY ANYTHING IN THIS CELL". Another large source of clones (about 4700 occurrences in total) are various licence texts. The most common license is Apache 2.0 (2889 occurrences).

The 2nd and 3rd most common clones, which do not seem to be intrinsically linked to a specific assignment or template, are snippets that consist solely of import state-ments. There are several recurring import-only snippets, both grouping the same libraries in different orders, or different groups of libraries. At 3475 occurrences, the most frequently occurring import-only snippet (of at least 4 lines) is this:

```
import pandas as pd
import numpy as np
import matplotlib.pyplot as plt
%matplotlib inline
```

The most common non-import only snippet that seems not to stem from a particular template or assignment is this function, which loads an image into a numpy array.

```
def load_image_into_numpy_array(image):
  (im_width, im_height) = image.size
  return np.array(image.getdata()).reshape(
    (im_height, im_width, 3)).astype(np.uint8)
```

**Single-Line Clones**  The most common single-line snippets are two imports 'import numpy as np' (63 049), 'import pandas as pd' (55 158), followed by the "magic function" '%matplotlib inline' (48 797) which controls the backend of matplotlib. The 4th and 5th most common snippets concern data frames: 'df.head()' (47 153) and 'df' (42 101). Finally, the 6th most common snippet prints the value of the variable 'a' (29 926). The 7th and 8th are imports of both numpy and pandas (26 059), respectively tensorflow (20 005). The 9th and 10th prints the variables 'x' (19 853) and 's' (19 065) respectively.

**Naming**  There is a proliferation of abbreviations of library names, and the abbreviations are relatively stable — pandas imported as *pd*, numpy as *np*, etc. A large number of data frames are called *df*, followed by *df1*, *df2*, etc. In the most common snippets, the most common variable names are *df*, *data* and *arr*. Also common are *s1*, *pwd*, *ls*, *my_list*, *ost*, *sc*, *list1* and *list2*. There is a surprising proliferation of single-letter variable names, most commonly *a*, *X*, *x*, *b*, *s*, *l*, *y*, *d*, *c*, *A*. These regularities might be contibuting to accidental clones.

## D  Artefacts

Our code used to compute the results presented in this paper can be found at

https://github.com/fxpl/notebooks

The dumping of Python notebooks (described in the methodology discussion of section 5.2) was done with the script run_pythonDumper.sh in commit 8bf959f3. The NotebookAnalyzer program was run using the script run_notebookAnalyzer.sh, commit f7574e50. To execute SccOutputAnalyzer, we used the script run_sccOutputAnalyzer.sh, commit 86ab409a. The post processing of the results from the analyzers was done as demonstrated in run_post_processing.sh, commit 4b657470

**About the authors**

**Malin Källén** is a PhD student in scientific computing at Uppsala University. She will defend her thesis in February.

**Ulf Sigvardsson** is a software professional. He worked on this project as part of his bachelor thesis.

**Tobias Wrigstad** is a Professor in Computer Science at Uppsala University, in Uppsala, Sweden. Contact him at: tobias.wrigstad@it.uu.se.